\documentclass[useAMS,usenatbib]{mn2e}

\usepackage[T1]{fontenc}
\usepackage{ae,aecompl}
\usepackage[dvips]{graphicx}
\usepackage{epsfig}
\usepackage{rotating}
\usepackage{graphicx}
\usepackage[caption=false]{subfig}
\usepackage{color}
\usepackage{hyperref}

\title[The accretion history of dark halos I]
{The accretion history of dark matter halos I:\\ The physical origin of the universal function}
\author[C.A.~Correa, J.S.B.~Wyithe, J.~Schaye and A.R.~Duffy]
  {Camila A.~Correa$^1$\thanks{E-mail: correac@student.unimelb.edu.au}, J. Stuart B.~Wyithe$^1$, Joop~Schaye$^2$ and Alan R.~Duffy$^{1,3}$\\
  $^1$ School of Physics, University of Melbourne, Parkville, VIC 3010, Australia\\
  $^2$ Leiden Observatory, Leiden University, PO Box 9513, 2300 RA Leiden, The Netherlands\\
  $^3$ Centre for Astrophysics and Supercomputing, Swinburne University of Technology, Melbourne, VIC 3122, Australia}
\date{\today}

\pagerange{\pageref{firstpage}--\pageref{lastpage}} \pubyear{2013}

\def\LaTeX{L\kern-.36em\raise.3ex\hbox{a}\kern-.15em
    T\kern-.1667em\lower.7ex\hbox{E}\kern-.125emX}

\begin{document}

\label{firstpage}

\maketitle

\begin{abstract}
Understanding the universal accretion history of dark matter halos is the first step towards determining the origin of their structure. We use the extended Press-Schechter formalism to derive the halo mass accretion history from the growth rate of initial density perturbations. We show that the halo mass history is well described by an exponential function of redshift in the high-redshift regime. However, in the low-redshift regime the mass history follows a power law because the growth of density perturbations is halted in the dark energy dominated era due to the accelerated expansion of the Universe. We provide an analytic model that follows the expression ${M(z)=M_{0}(1+z)^{af(M_{0})}e^{-f(M_{0})z}}$, where ${M_{0}=M(z=0)}$, $a$ depends on cosmology and $f(M_{0})$ depends only on the linear matter power spectrum. The analytic model does not rely on calibration against numerical simulations and is suitable for any cosmology. We compare our model with the latest empirical models for the mass accretion history in the literature and find very good agreement. We provide numerical routines for the model online\footnote{}.
\end{abstract}


\footnotetext{Available at \href{https://bitbucket.org/astroduff/commah}{\it{https://bitbucket.org/astroduff/commah}}} 

\section{Introduction}

Throughout the last decade, there have been many attempts to quantify halo mass accretion histories using catalogues of halos from numerical simulations (\citealt{Wechsler,McBride,Wang,Genel,Fakhouri,Voort,Faucher-Giguere, Johansson,Benson,Wu,Behroozi}). \citet{Wechsler} characterized the mass history of halos more massive than $10^{12} \rm{M}_{\sun}$ at $z=0$ using a one-parameter exponential form $e^{\beta z}$. In their work, \citet{Wechsler} limited their analysis to the build-up of clusters through progenitors already larger than the Milky Way halo. Similarly, \citet{McBride} limited their analysis to massive halos and found that a large fraction were better fitted when an additional factor of $(1+z)^{\alpha}$ was added to the \citet{Wechsler} exponential parametrization, yielding a mass history of the form $M\propto (1+z)^{\alpha}e^{\beta z}$. \citet{Wong} investigated whether the mass history can be described by a single parameter function or whether more variables are required. They utilized principal component analysis and found that despite the fact that the \citet{McBride} two-parameter formula presents an excellent fit to halo mass histories, the parameters $\alpha$ and $\beta$ are not a natural choice of variables as they are strongly correlated. Recently, \cite{vandenBosch14} studied halo mass histories extracted from $N$-body simulations and semi-analytical merger trees. However, so far no universal and physically motivated model of a universal halo mass history function has been provided.

An alternative method to interpret the complex numerical results and to unravel the physics behind halo mass growth, is the extended Press Schechter (EPS) formalism. EPS theory provides a framework that allows us to connect the halo mass accretion history to the initial density perturbations. \citet{Neistein} showed in their work that it is possible to create halo mass histories directly from EPS formalism by deriving a useful analytic approximation for the average halo mass growth. In this work we aim to provide a physical explanation for the `shape' of the halo mass history using the EPS theory and the analytic formulation of \citet{Neistein}. The resulting model for the halo mass history, which is suitable for any cosmology, depends mainly on the linear power spectrum.

This paper is organized as follows. We show in Section \ref{MAHwithEPS} that the halo mass history is naturally described by a power law and an exponential as originally suggested from fits to cosmological simulation data by \citet{McBride}. We then provide a simple {\it{analytic model}} based on the EPS formalism and compare it to the latest empirical halo mass history models from the literature. Finally, we provide a summary of formulae and discuss our main findings in Section \ref{SummaryAndConclussion}. 

In a companion paper, \citet{PaperII} hereafter Paper II, we explore the relation between the structure of the inner dark matter halo and halo mass history using a suite of cosmological simulations. We provide a semi-analytic model for halo mass history that combines analytic relations for the concentration and formation time with fits to simulations, to relate halo structure to the mass accretion history. This semi-analytic model has the functional form, $M=M_{0}(1+z)^{\alpha}e^{\beta z}$, where the parameters $\alpha$ and $\beta$ are directly correlated with the dark matter halo concentration. Finally, in a forthcoming paper (\citealt{PaperIII}, hereafter Paper III), we combine the semi-analytic model of halo mass history with the analytic model described in Section \ref{MAHmodel} of this paper to predict the concentration-mass relation and its dependence on cosmology.

\section{Analytic model for the halo mass history}\label{MAHwithEPS}

In order to provide a physical motivation for the `shape' of the halo mass history, we begin in Section 2.1 with an analytic study of dark matter halo growth using the EPS formalism (\citealt{Bond,Lacey}). In Section 2.2, we show that the halo mass history is well described by an exponential at high redshift, and by a power law at low redshift. In Section 2.3, we then adopt the power-law exponential form and use it to provide a simple analytic model for halo mass histories. Finally, we compare our results with the latest models of halo mass history from the literature in Section 2.4.

\subsection{Theoretical background of EPS theory}

The EPS formalism is an extension of the Press-Schechter (PS) formalism (\citealt{Press}), which provides an approximate description of the statistics of merger trees using a stochastic process. The EPS formalism has been widely used in algorithms for the construction of random realizations of merger trees (\citealt{Kauffmann,Benson05,Cole08,Neistein08}). 

In the standard model of cosmology, the structures observed today are assumed to have grown from small initial density perturbations due to the action of gravity. The initial density contrast, defined as $\delta({\bf{x}},t)=\rho({\bf{x}},t)/\bar{\rho}-1$ (with $\rho({\bf{x}},t)$ density field and $\bar{\rho}$ mean density), is considered to be a Gaussian random field completely specified by the power spectrum $P(k)$, where $k$ is a spatial frequency. In the linear regime ($\delta\ll 1$), the perturbations determined by $\delta({\bf{x}},t)$ evolve as $\delta({\bf{x}},t)=\delta_{0}({\bf{x}})D(t)$, where $\delta_{0}({\bf{x}})$ is the density contrast field linearly extrapolated to the present time and $D(t)$ is the linear growth factor. According to the spherical collapse model, once $\delta({\bf{x}},t)$ exceeds a critical threshold $\delta_{\rm{crit}}^{0}\simeq 1.69$ (with a weak dependence on redshift and cosmology) the perturbation starts to collapse. Regions that have collapsed to form a virialized object at redshift $z$ are then associated with those regions for which 

\begin{equation}\label{delta}
\delta_{0}>\delta_{\rm{c}}(z)\equiv \frac{\delta_{\rm{crit}}^{0}}{D(z)}=\frac{1.686\Omega_{\rm{m}}(z)^{0.0055}}{D(z)}.
\end{equation}

\noindent Here $\Omega_{\rm{m}}(z)=\Omega_{\rm{m},0}(1+z)^{3}H_{0}^{2}/H(z)^{2}$, $H(z)=H_{0}[\Omega_{\rm{m},0}(1+z)^{3}+\Omega_{\Lambda,0}]^{1/2}$ and the linear growth factor $D(z)$ is computed by performing the integral

\begin{equation}\label{Dzeq}
D(z)\propto H(z)\int_{z}^{\infty}\frac{1+z'}{H(z')^{3}}dz',
\end{equation}

\noindent where $D(z)$ is normalized to unity at the present day. The collapsed regions are assigned masses by smoothing the density contrast $\delta_{0}$ with a spatial window function. In what follows, instead of using the halo mass as the independent variable, we adopt the variance of the smoothed density field, $\sigma^{2}(M)$.

The EPS model developed by \citet{Bond} is based on the excursion set formalism. For each collapsed region one constructs random `trajectories' of the linear density contrast $\delta(M)$ as a function of the variance $\sigma^{2}(M)$. Defining

\begin{equation}\nonumber
\omega\equiv \delta_{\rm{c}}(z)\quad{\rm{and}}\quad S\equiv\sigma^{2}(M),
\end{equation}

\noindent we use $\omega$ and $S$ to label redshift and mass, respectively. If the initial density field is a Gaussian random field smoothed under a sharp {\it{k}}-space filter, increasing $S$ (corresponding to a decreased in the filter mass $M$) results in $\delta(M)$ starting to wander away from zero, executing a random walk. The fraction of matter in collapsed objects in the mass interval $M,M+dM$ at redshift $z$ is associated with the fraction of trajectories that have their first upcrossing through the barrier $\omega$ in the interval $S,S+dS$, which is given by (\citealt{Bond,Bower,Lacey}),

\begin{equation}\label{prob1}
f(S,\omega)dS=\frac{1}{\sqrt{2\pi}}\frac{\omega}{S^{3/2}}\rm{exp}\left[-\frac{\omega^{2}}{2S}\right]dS,
\end{equation}

\noindent where $S$ is defined as

\begin{equation}\label{S}
S(M) = \frac{1}{2\pi^{2}}\int_{0}^{\infty}P(k)\hat{W}^{2}(k;R)k^{2}dk.
\end{equation}

\noindent Here $P(k)$ is the linear power spectrum and $\hat{W}(k;R)$ is the Fourier transform of a top hat window function. The probability function in eq.~(\ref{prob1}) yields the PS mass function and gives the probability for a change $\Delta S$ in a time step $\Delta\omega$, since for random walks the upcrossing probabilities are a Markov process (i.e. are independent of the path taken). The analytic function given by eq.~(\ref{prob1}) provides the basis for the construction of merger trees. 
\citet{Neistein} derived a differential equation for the average halo mass history over an ensemble of merger trees from the EPS formalism. Defining $M_{\rm{EPS}}(z)$ to be the mass of the most massive halo (main progenitor), along the main branch of the merger tree, as a function of redshift, they obtained the differential equation (see derivation in appendix A)   

\begin{eqnarray}\label{dMdz_EPS}
\frac{dM_{\rm{EPS}}}{dz}&=& \sqrt{\frac{2}{\pi}}\frac{M_{\rm{EPS}}}{\sqrt{S_{q}-S}}\frac{1.686}{D(z)^{2}}\frac{dD(z)}{dz},
\end{eqnarray}

\noindent where $S_{q}=S(M_{\rm{EPS}}(z)/q)$ and $S=S(M_{\rm{EPS}}(z))$. The value of $q$ needs to be obtained empirically so that $M_{\rm{EPS}}$ reproduces halo mass histories from cosmological simulations. \citet{Neistein} showed that the uncertainty of $q$ is an intrinsic property of EPS theory, where different algorithms for constructing merger trees may correspond to different values of $q$. 

\begin{figure}
\begin{center}
\includegraphics[width=0.45\textwidth]{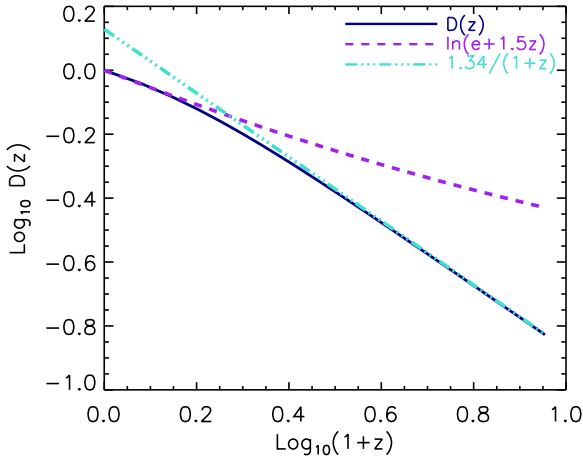}
\caption{Linear growth factor against redshift. The dark blue solid line shows the growth factor obtained by performing the integral given by eq.~(\ref{Dzeq}). The purple dashed line corresponds to the low-redshift approximation in eq.~(\ref{Dz}). Similarly, the green dot-dashed line shows the approximation of the growth factor in the high-redshift regime.}
\label{Dzfig}
\end{center}
\end{figure}

\subsection{Mass accretion in the high- and low-$z$ regimes}\label{Motivation}

In this section we analyse how the evolution of $M_{\rm{EPS}}(z)$ given by eq.~(\ref{dMdz_EPS}) is governed by the growth factor. We provide two practical approximations for the growth factor in the high- and low-redshift regimes and investigate the `shape' of $M_{\rm{EPS}}(z)$ in these regimes by integrating eq.~(\ref{dMdz_EPS}). 

In addition to the redshift dependence of the growth factor ($D(z)$), in eq.~(\ref{dMdz_EPS}) an extra redshift dependence is introduced through the quantity $[S_{q}-S]^{-1/2}=[S(M(z)/q)-S(M(z))]^{-1/2}$. Before integrating eq. (\ref{dMdz_EPS}), we calculate how the value of $[S(M(z)/q)-S(M(z))]^{-1/2}$ changes with redshift to find a suitable first order approximation to simplify the calculations. 

We replace ${S=\sigma^{2}}$ and approximate ${\sigma\approx M^{\gamma}}$, where $\gamma=-0.063$ for ${M\le 10^{12}\rm{M}_{\sun}}$ and $\gamma=-0.21$ for $M>10^{12}\rm{M}_{\sun}$, to obtain

\begin{eqnarray}\label{approx}
\frac{[S(M(z)/q)-S(M(z))]^{-1/2}}{[S(M_{0}/q)-S(M_{0})]^{-1/2}} &=&\left(\frac{M(z)}{M_{0}}\right)^{-\gamma},
\end{eqnarray}

\noindent where ${M_{0}=M(z=0)}$. 

Given the weak dependence on mass in the right part of eq.~(\ref{approx}), we simplify the expression $[S(M(z)/q)-S(M(z))]^{-1/2}$ in our analysis using the approximation $[S(M(z)/q)-S(M(z))]^{-1/2}\approx [S(M_{0}/q)-S(M_{0})]^{-1/2}$. It is important to note that as the $M_{\rm{EPS}}(z)$ evolution (given by eq.~\ref{dMdz_EPS}) is only governed by the growth factor in eq.~(\ref{Dzeq}), we find that in the latter part of the accretion history, where most mass is accreted, the approximation $[S(M(z)/q)-S(M(z))]^{-1/2}\approx [S(M_{0}/q)-S(M_{0})]^{-1/2}$, will carry an $\sim 5\%$ error for $M(z)\le 10^{12}\rm{M}_{\sun}$ and $15\%$ error for $M(z)> 10^{12}\rm{M}_{\sun}$. Earlier in the accretion history, the errors may be as large as $\sim 20\%$ and $40\%$ for $M(z)\le 10^{12}\rm{M}_{\sun}$ and $M(z)> 10^{12}\rm{M}_{\sun}$, respectively. We demonstrate in Section \ref{comparison_sec} that these errors do not affect the final $M(z)$ model, which we show provides very good agreement with simulation-based mass history models from the literature.

The growth factor can be approximated with high accuracy by 

\begin{equation}\label{Dz}
D(z) = \left\{
\begin{array}{cl}
\frac{1.34}{1+z} & \rm{if}\quad z \gg 1,\\
\\\nonumber
\frac{1}{\ln(e+1.5z)} & \rm{if}\quad z \ll 1,
\end{array} \right.
\end{equation}

\noindent for all cosmologies. Fig. \ref{Dzfig} shows the growth factor as given by eq. (\ref{Dzeq}) (solid dark blue line), together with the approximations in the high- and low-redshift regimes (dot-dashed green and purple lines, respectively). The high-redshift approximation for $D(z)$ is an exact solution for an Einstein-de Sitter (EdS) cosmology ($\Omega_{\Lambda}=0$). However, the growth rate slows down in the cosmological constant dominated phase, so that linear perturbations grow faster in an EdS universe.

We can estimate $M_{\rm{EPS}}(z)$ in the high- and low-redshift regimes by substituting the two expressions from eq. (\ref{Dz}) into eq.~(\ref{dMdz_EPS}). In the high-redshift regime (where $z\gg 1$), 

\begin{eqnarray}\label{approx}
\frac{dM_{\rm{EPS}}}{M_{\rm{EPS}}} &=& \sqrt{\frac{2}{\pi}}\frac{1}{\sqrt{S_{q}-S}}\frac{1.686}{D(z)^{2}}\frac{dD(z)}{dz}dz,\\\nonumber
\frac{dM_{\rm{EPS}}}{M_{\rm{EPS}}} & = & -f(M_{0})dz,\nonumber
\end{eqnarray}

\noindent where $f(M_{0})=1/\sqrt{S(M_{0}/q)-S(M_{0})}$ is a function of halo mass. Integrating this last equation, we obtain

\begin{eqnarray}\label{highz}
M_{\rm{EPS}}(z) &=& M_{0}e^{-f(M_{0})z}\quad \rm{at} \quad z\gg 1.
\end{eqnarray}

\noindent Thus, we conclude that the halo mass history is well described by an exponential ($M(z)\sim e^{\beta z}$) in the high-redshift regime, as suggested by \citet{Wechsler}. In the low-redshift regime, from eq. (\ref{approx}) and the bottom part of eq.~(\ref{Dz}), we find

\begin{eqnarray}\nonumber
\frac{dM_{\rm{EPS}}}{M_{\rm{EPS}}} & = & -\frac{1.34}{1.8+z}f(M_{0})dz.\nonumber
\end{eqnarray}

\noindent Integrating the above expression yields

\begin{equation}\label{lowz}
M_{\rm{EPS}}(z)= M_{0}(1+0.5z)^{-1.34f(M_{0})}\quad \rm{at} \quad z\ll 1.
\end{equation}

Therefore, in the low-redshift regime, a power law ($M(z)\sim(1+z)^{\alpha}$) is necessary because the growth of density perturbations is halted in the dark energy dominated era due to the accelerated expansion of the Universe. 

\begin{figure}
\begin{center}
\includegraphics[width=0.48\textwidth]{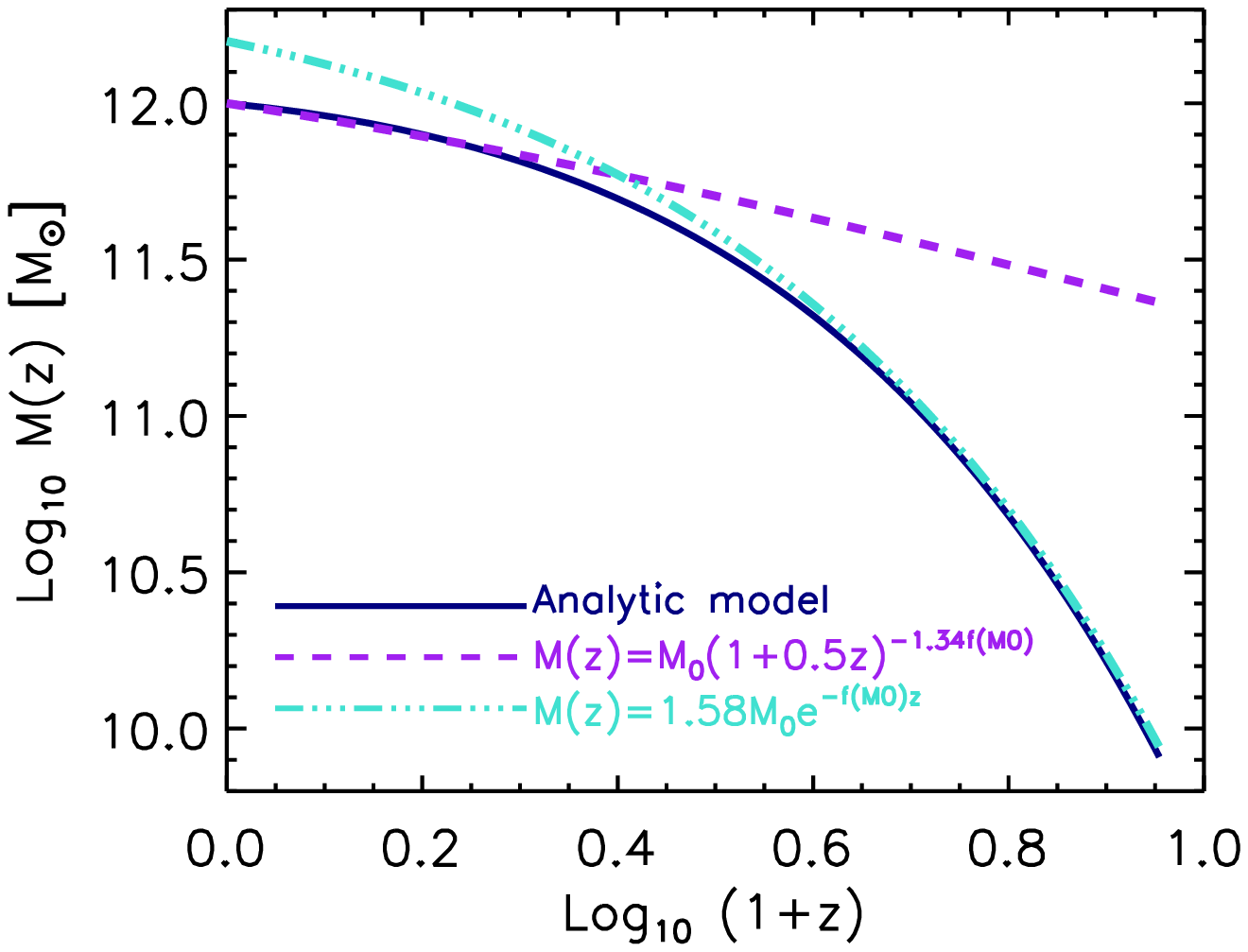}\\
\includegraphics[width=0.48\textwidth]{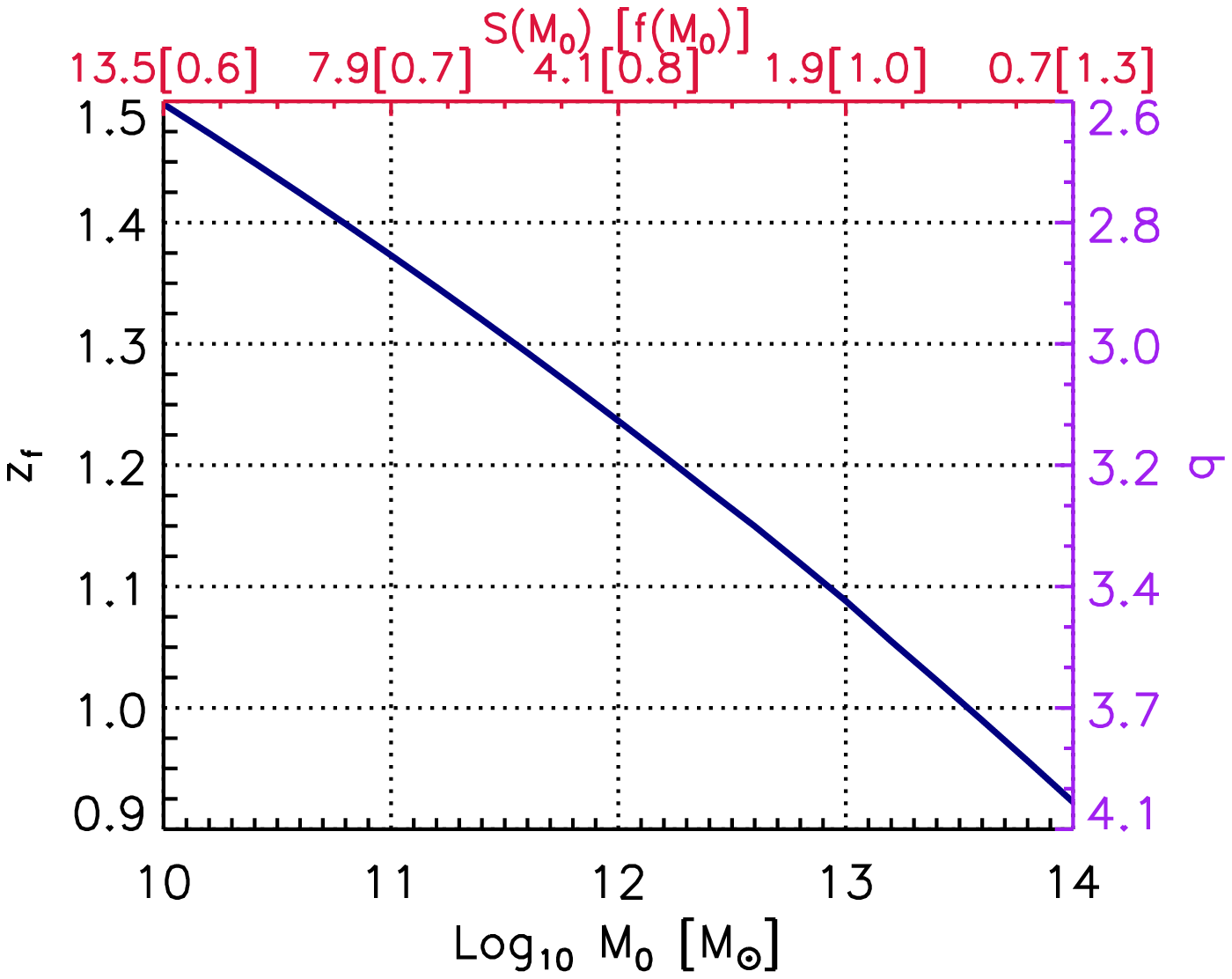}
\caption{{\it{Top panel}}: comparison between halo mass histories predicted by the analytic model $M(z)=M_{0}(1+z)^{af(M_{0})}e^{-f(M_{0})z}$, given by eqs. (\ref{analytic_model_1})-(\ref{analytic_model_5}) (blue solid line), and the approximated halo mass histories given by eq.~(\ref{lowz}), for the low-redshift regime (purple dashed line), and by eq.~(\ref{highz}), for the high-redshift regime (green dot-dashed line). {\it{Bottom panel}}: formation redshift against halo mass. Here the formation redshifts were obtained by solving eqs. (\ref{zf_1eq}) and (\ref{zf_2eq}). The right Y-axis shows the values of `$q$' obtained when calculating the formation redshift, whereas the top X-axis shows the variance of the smoothed density field of a region that encloses the mass  indicated by the bottom X-axis. The values of the function $f(M_{0})=1/\sqrt{S(M_{0}/q)-S(M_{0})}$ are shown in brackets.}
\label{MzEPS_approx}
\end{center}
\end{figure}

\subsection{Analytic mass accretion history model based on the EPS formalism}\label{MAHmodel}

In this subsection we provide an analytic model for the halo mass history based on the EPS formalism. This model is not calibrated against  numerical simulations and allows an exploration of the physical processes involved in the halo mass growth. Based on our analysis above, we begin by assuming that the halo mass history is well described by the simple form

\begin{equation}\label{Mz_expr}
M(z)=M_{0}(1+z)^{\alpha}e^{\beta z}.
\end{equation} 

\noindent The presence of the function $S(M_{0})$ in both the high-redshift exponential and the low-redshift power law explains the correlation between $\alpha$ and $\beta$ found by \citet{Wong}. We estimate the relation between $\alpha$, $\beta$ and $S(M_{0})$ by replacing $M_{\rm{EPS}}(z)$ in eq.~(\ref{dMdz_EPS}) with the above expression. Evaluating at $z=0$ we obtain

\begin{equation}
\alpha+\beta = 1.686(2/\pi)^{1/2}f(M_{0})\frac{dD}{dz}|_{z=0}.
\end{equation}

\noindent Assuming $\beta$ follows the relation with $S(M_{0})$ shown in eq.~(\ref{highz}) for the high-$z$ regime we find 
\begin{eqnarray}
\beta &=& -f(M_{0}),\\
\alpha &=& af(M_{0}),
\end{eqnarray}

\noindent with $a=\left[1.686(2/\pi)^{1/2}\frac{dD}{dz}|_{z=0}+1\right]$. The above equations introduce a halo mass history model directly derived from the EPS theory, where the parameters $\alpha$ and $\beta$ are related through the variance of the smoothed density field, $S(M_{0})$. The quantity $q$ is a free parameter which can be determined by adding an extra equation that restricts the model. We do this by defining the halo formation redshift, $\tilde{z}_{\rm{f}}$, as the redshift where $M(\tilde{z}_{\rm{f}})=M_{0}/q$. From eq. (\ref{Mz_expr}) we obtain

\begin{equation}\label{zf_1eq}
\frac{1}{q} = (1+\tilde{z}_{\rm{f}})^{af(M_{0})}e^{-f(M_{0})\tilde{z}_{\rm{f}}},
\end{equation}  

\noindent where $f(M_{0})=1/\sqrt{S(M_{0}/q)-S(M_{0})}$.

The general relation between formation time and $q$ was introduced by \citet{Lacey}, using the expression

\begin{equation}\label{Mz_expr_EPS}
M(z)=M_{0}\left[1-{\rm{erf}}\left(\frac{\delta_{\rm{c}}(z)-\delta_{\rm{c}}(0)}{\sqrt{2(S(M_{0}/q)-S(M_{0})}}\right)\right],
\end{equation} 

\noindent which describes the average mass of the main progenitors in the EPS merger tree. We use eq.~(\ref{Mz_expr_EPS}) to evaluate the halo mass at $\tilde{z}_{\rm{f}}$ and find the distribution of formation times. We follow the approach of \citet{Lacey} and find

\begin{equation}\label{zf_2eq}
\delta_{c}(\tilde{z}_{\rm{f}})=\delta_{\rm{c}}(0)+\sqrt{2}f^{-1}(M_{0}){\rm{erf}}^{-1}(1-1/q).
\end{equation}

\noindent We solve eqs. (\ref{zf_1eq}) and (\ref{zf_2eq}) and find $q$ and $\tilde{z}_{\rm{f}}$ for various halo masses. We then fit the $q-M_{0}$ and $\tilde{z}_{\rm{f}}-q$ relations using a second order polynomial in $\log_{10}M$ for $\tilde{z}_{\rm{f}}$, and obtain the following set of equations that describe the halo mass history,

\begin{eqnarray}\label{analytic_model_1}
M(z)&=&M_{0}(1+z)^{af(M_{0})}e^{-f(M_{0})z},\\\label{analytic_model_2}
a&=&\left[1.686(2/\pi)^{1/2}\frac{dD}{dz}|_{z=0}+1\right],\\\label{analytic_model_3}
f(M_{0})&=&1/\sqrt{S(M_{0}/q)-S(M_{0})},\\\label{analytic_model_4}
q&=&4.137\tilde{z}^{-0.9476}_{\rm{f}},\\\label{analytic_model_5}
\tilde{z}_{\rm{f}}&=&-0.0064(\log_{10}M_{0})^{2}+0.0237(\log_{10}M_{0})\\\nonumber
&& +1.8837.
\end{eqnarray}

\noindent The equations that relate $q$, $\tilde{z}_{\rm{f}}$ and $M_{0}$ are calculated assuming the WMAP5 cosmology, but work for others cosmologies\footnote{We verified that the MAHs predicted by eqs. (\ref{analytic_model_1}-\ref{analytic_model_5}) are in excellent agreement with the simulations for the WMAP1/3/9 and Planck cosmologies.} because the halo mass histories are mainly driven by the change in $\sigma_{8}$ and $\Omega_{\rm{m}}$. We reiterate that unlike previous models based on EPS theory (e.g. \citealt{van}), the analytic model specified in the above equations was not calibrated against any simulation data. 

The top panel of Fig.~\ref{MzEPS_approx} shows a comparison between the analytic model given by eqs. (\ref{analytic_model_1})-(\ref{analytic_model_5}) (blue solid line), and the limiting case for the halo mass histories given by eq.~(\ref{lowz}) for the low-redshift regime (purple dashed line) and by eq.~(\ref{highz}) for the high-redshift regime (green dot-dashed line). In the last case, we renormalized the mass history curve to match that given by the analytic model at $z=7$. This figure demonstrates how exponential growth dominates the mass history at high redshift, and power-law growth dominates at low redshift, as concluded in the previous section.

The bottom panel of Fig. \ref{MzEPS_approx} shows the formation time obtained from eqs. (\ref{zf_1eq}) and (\ref{zf_2eq}), as a function of halo mass. As expected, larger mass halos form later. The right Y-axis shows the values of $q$ obtained when calculating formation time, whereas the top X-axis shows the variance of the smoothed density field of a region that encloses the mass indicated by the bottom X-axis. The values of the function $f(M_{0})=1/\sqrt{S(M_{0}/q)-S(M_{0})}$ are included in brackets. As can be seen from this figure, the larger the halo mass, the lower the variance $S(M_{0})$, the larger $f(M_{0})$, and so the larger the factor in the exponential that makes the halo mass halt its rapid growth at low redshift. For example, a $10^{14}\rm{M}_{\sun}$ halo has a mass history mostly characterized by an exponential growth ($\sim e^{-f(M_{0})z}$) until redshift $z=1/f(M_{0})=0.7$, whereas a $10^{10}\rm{M}_{\sun}$ halo only has an exponential growth until redshift $z=1.6$.  Note, however, that our analytic model is not limited to the halo mass ranges shown in the bottom panel of Fig.~\ref{MzEPS_approx}, it can be extended to any halo masses and redshifts and the $q-M_{0}$ and $\tilde{z}_{\rm{f}}-M_{0}$ relations still hold. 

In addition to the halo mass history, it is possible to calculate the accretion rate of a halo at a particular redshift. In order to do that we differentiate eq.~(\ref{Mz_expr}) with respect to time and replace $dz/dt$ by $-H_{0}[\Omega_{\rm{m}}(1+z)^{5}+\Omega_{\Lambda}(1+z)^{2}]^{1/2}$ to obtain

\begin{eqnarray}\nonumber
\frac{dM(z)}{dt} &=& 71.6{\rm{M}}_{\sun}{\rm{yr}}^{-1} \left(\frac{M(z)}{10^{12}\rm{M}_{\sun}}\right)\left(\frac{h}{0.7}\right) \\\nonumber
&& \times f(M_{0})[(1+z)-a][\Omega_{\rm{m}}(1+z)^{3}+\Omega_{\Lambda}]^{1/2},
\end{eqnarray}

\noindent where $a$ is given by eq.~(\ref{analytic_model_2}) and $f(M_{0})$ is given by eq.~(\ref{analytic_model_3}). Note that the above formula will give the accretion rate at redshift $z$ of a halo that has mass $M_{0}$ at redshift $z=0$, and mass $M(z)$ at redshift $z$.

The physical relation derived between the parameters describing the exponential and power law behaviour implies that a single parameter accretion history formula should be seen in numerical simulations. In Paper II we investigate the $\alpha$ and $\beta$ parameter dependence in more detail, and we determine the intrinsic relation (which cannot be explored under the EPS formalism) between halo assembly history and inner halo structure. 

\subsection{Comparison with previous studies}\label{comparison_sec}

\begin{figure}
\begin{center}
\includegraphics[width=0.4\textwidth]{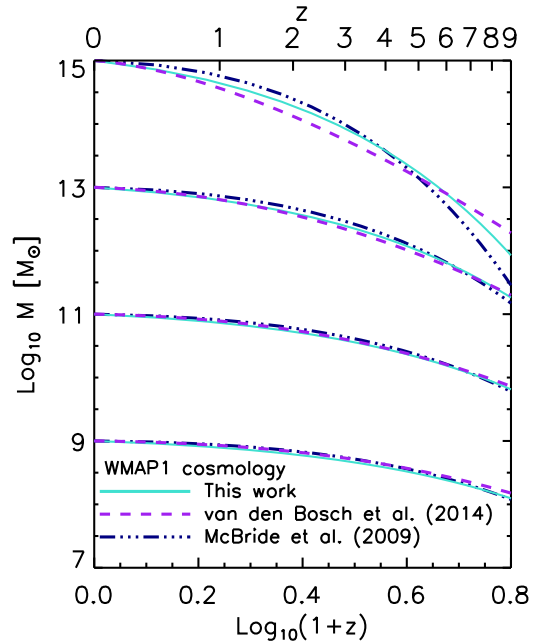}
\caption{Comparison of halo mass history models. The analytic model presented in this work (turquoise solid lines) is compared with the median mass history obtained from the Bolshoi simulation and merger trees from \citet{vandenBosch14} (purple dashed lines) and the best-fitting relations from the Millennium simulation from \citet{McBride} (dark blue dot dashed lines). The comparisons are shown for four halo masses and for consistency with \citet{McBride} we assumed in our model and in the \citet{vandenBosch14} model the WMAP1 cosmology.}
\label{Mz_comparison}
\end{center}
\end{figure}

In this section we briefly describe the simulation-based halo mass history models presented in \citet{vandenBosch14} (vdB14) and \citet{McBride} (MB09), and contrast them with our analytic model given by eqs. (\ref{analytic_model_1}-\ref{analytic_model_5}). Fig.~\ref{Mz_comparison} shows a comparison of our mass history model (turquoise solid lines) to the models of vdB14 (purple dashed lines) and MB09 (dark blue dot dashed lines). vdB14 used the mass histories from the Bolshoi simulation (\citealt{Klypin}) and extrapolated them below the resolution limit using EPS merger trees. They then used a semi-analytic model to transform the average or median mass accretion history for a halo of a particular mass taken from the Bolshoi simulation, to another cosmology, via a simple transformation of the time coordinate. Using their publicly available code we calculated the mass histories of $10^{9}$, $10^{11}$, $10^{13}$ and $10^{15}\rm{M}_{\sun}$ halos for the WMAP1 cosmology. We find good agreement between our model and vbB14 for all halo masses. The main difference occurs for the mass histories of high-mass halos ($M_{0}>10^{13}\rm{M}_{\sun}$), where vdB14 seems to underpredict the mass growth above $z=4$ by a factor of $\sim 1.2$. In addition, vdB14 compared their model to those of \citet{Zhao09} and \citet{Giocoli}, and found that both works predict smaller halo mass growth at $z>1.5$.

We also compare our model to the MB09 mass history curves. MB09 used the Millennium simulation (\citealt{Springel05}) and separated their halo sample into categories depending on the `shape' of the mass histories, from late-time growth that is steeper than exponential to shallow growth. We find that the fitting function that best matches our results is from their type IV category. We find good agreement with the MB09 formula.

Fig.~\ref{Mz_comparison} demonstrates that the physically motivated analytic model presented in this work yields mass histories that are in good agreement with the results obtained from numerical simulations. However, in contrast to the models based on fits to simulation results, our analytic model can be extrapolated to very low masses and is suitable for any cosmology.

\section{Summary and conclusion}\label{SummaryAndConclussion}

In this work we have demonstrated how halo mass histories are determined by the initial power spectrum of density fluctuations, and the growth factor. We found that the halo mass history is well described by an exponential ($M(z)\sim e^{\beta z}$, as suggested by \citealt{Wechsler}) in the high-redshift regime, but that the accretion slows to a power law at low redshift ($M(z)\sim(1+z)^{\alpha}$) because the growth of density perturbations is halted in the dark energy dominated era due to the accelerated expansion of the Universe. The resulting expression

\begin{equation}\nonumber
M(z)= M_{0}(1+z)^{\alpha}e^{\beta z},
\end{equation}

\noindent accurately captures all halo mass histories (Fig. \ref{Mz_comparison}). Adopting this expression, we provided an analytic mass history model based on the EPS formalism, in which the parameters $\alpha$ and $\beta$ are related to the power spectrum by

\begin{eqnarray}\nonumber
\beta &=& -f(M_{0}),\\\nonumber
\alpha &=& \left[1.686(2/\pi)^{1/2}\frac{dD}{dz}|_{z=0}+1\right]f(M_{0}),\\\nonumber
f(M_{0}) &=& [S(M_{0}/q)-S(M_{0})]^{-1/2},\\\nonumber
S(M) &=& \frac{1}{2\pi^{2}}\int_{0}^{\infty}P(k)\hat{W}^{2}(k;R)k^{2}dk,\nonumber
\end{eqnarray} 

\noindent where $D$ is the linear growth factor, $P$ is the linear power spectrum, and $q$ is related to the total halo mass as

\begin{eqnarray}\nonumber
q &=& 4.137\tilde{z}^{-0.9476}_{\rm{f}},\\\nonumber
\tilde{z}_{\rm{f}} &=& -0.0064(\log_{10}M_{0})^{2}+0.0237(\log_{10}M_{0})\\\nonumber
&& +1.8837.
\end{eqnarray}

We found very good agreement between the halo mass histories predicted by our analytic model and published fits to simulation results (Fig. \ref{Mz_comparison}). The reader may find a step-by-step description on how to implement both models in Appendix \ref{MAH_Description}, as well as numerical routines online\footnote{Available at \href{https://bitbucket.org/astroduff/commah}{\it{https://bitbucket.org/astroduff/commah}}.}.

The relation of the parameters $\alpha$ and $\beta$ with the linear power spectrum explains the correlation between the dark matter halo concentration and the linear {\it{rms}} fluctuation of the primordial density field that was previously noted in numerical simulations (\citealt{Prada,Diemer14}). We show it in Paper II, where we derive a semi-analytic model for the halo mass history that relates halo structure to the mass accretion history. In that work we combine the semi-analytic model with the analytic model presented here to establish the physical link between halo concentrations and the initial density perturbation field. Finally, in Paper III we combine the analytic and semi-analytic description to predict the concentration-mass relation of halos and its dependence on cosmology.
 

\section*{Acknowledgements}

We are grateful to the referee Aaron Ludlow for fruitful comments that substantially improved the original manuscript. CAC acknowledges the support of the 2013 John Hodgson Scholarship and the hospitality of Leiden Observatory. JSBW is supported by an Australian Research Council Laureate Fellowship. JS acknowledges support by the European Research Council under the European  Union's Seventh Framework Programme (FP7/2007-2013)/ERC Grant agreement 278594-GasAroundGalaxies. We are grateful to the OWLS team for their help with the simulations. 

\bibliography{biblio}
\bibliographystyle{mn2e}

\appendix

\appendix

\section{Differential equation for $M_{\rm{EPS}}$}

To construct the mass history of a given parent halo mass, it is most convenient to begin from the parent halo and go backwards in time following the merger events of the most massive progenitor. We begin by assuming that a halo of mass $M_{j}$ (corresponding to a mass variance $S_{j}$) at time $\omega_{j}$ takes a small time-step $\Delta\omega$ back in time (note that $\Delta\omega<0$). At the time $\omega_{j+1}=\omega_{j}+\Delta\omega$, we calculate the average mass of the main progenitor $M_{j+1}$ ($S_{j+1}$) following an excursion set approach and computing the probability for a random walk originating at $(S_{j},\omega_{j})$ and executing a first upcrossing of the barrier $\omega_{j+1}$ at $S_{j+1}$. Hence the probability we want is given by eq.~(\ref{prob1}) upon replacing $S$ by $S_{j+1}-S_{j}$ and $\omega$ by $\omega_{j+1}-\omega_{j}$. Converting from mass weighting to number weighting, one obtains the average number of progenitors at $z_{j+1}$ in the mass interval ($M_{j+1},M_{j+1}+dM$) which by time $z_{j}$ have merged to form a halo of mass $M_{j}$,

\begin{eqnarray}\label{prob2}
P(M_{j+1},z_{j+1}|M_{j},z_{j})dM_{j+1}=\frac{M_{j}}{M_{j+1}}\\\nonumber
\times f(S_{j+1},\omega_{j+1}|S_{j},\omega_{j})\left\vert\frac{dS_{j+1}}{dM_{j+1}}\right\vert dM_{j+1}.
\end{eqnarray}

As a first approximation, we assume that $P(M_{j+1},z_{j+1}|M_{j},z_{j})=0$ for $M_{j+1}<M_{j}/q$, so that the main progenitor always has a mass $M_{j+1}\ge M_{j}/q$ for a given $q$ value. Therefore, the average mass of the main progenitor can be written as

$$M_{j+1}(z_{j+1})=\int_{M_{j}/q}^{M_{j}}P(M|M_{j},z_{j})MdM.$$

\noindent We then replace $P(M|M_{j},z_{j})$ by eq.~(\ref{prob2}) and integrate

$$M_{j+1}(z_{j+1})=\int_{M_{j}/q}^{M_{j}}\frac{M_{j}}{M}\frac{1}{\sqrt{2\pi}}\frac{\Delta\omega}{\Delta S^{3/2}}\rm{exp}\left[-\frac{\Delta\omega^{2}}{2\Delta S}\right]\left\vert\frac{dS}{dM}\right\vert MdM,$$

\noindent where $\Delta\omega = w_{j+1}-w_{j}$, that correspond to the redshift interval ($z_{j+1},z_{j}$), and $\Delta S=S-S_{j}$. Defining $u^{2}=\Delta\omega^{2}/2\Delta S$, the above integral yields

$$M_{j+1}(z_{j+1})=\frac{2}{\sqrt{\pi}}M_{j}\int_{u_{j+1}}^{u_{j}}e^{-u^{2}}du,$$

\noindent here $u_{j}=\Delta\omega/\sqrt{2(S_{j}-S_{j})}\rightarrow\infty$ and $u_{j+1}=\Delta\omega/\sqrt{2(S(M_{j}/q)-S_{j})}$. Therefore the integral can be written in terms of the error function

\begin{equation}\label{Meps}
M_{j+1}(z_{j+1})=M_{j}\left[1-\rm{erf}\left(\frac{\Delta\omega}{\sqrt{2S(M_{j}/q)-2S_{j}}}\right)\right].
\end{equation}

\noindent This equation becomes linear in $\Delta\omega$ for small enough $\Delta\omega$, therefore the mass history can be constructed by iterating

$$M_{\rm{EPS}}(\Delta\omega_{i}+\Delta\omega_{j}|M_{0})=M_{\rm{EPS}}(\Delta\omega_{i}|M_{\rm{EPS}}(\Delta\omega_{j}|M_{0})),$$

\noindent where we term $M_{0}$ as the mass of the parent halo. Then the rate of change, $dM_{\rm{EPS}}/d\omega$, can be computed as

\begin{eqnarray}\nonumber
\frac{dM_{\rm{EPS}}}{d\omega} &=& \lim_{\Delta\omega\rightarrow 0}\frac{M_{\rm{EPS}}(\Delta\omega)-M_{0}}{\Delta\omega},\\\nonumber
&=& -M_{0}\lim_{\Delta\omega\rightarrow 0}\frac{1}{\Delta\omega}\rm{erf}\left(\frac{\Delta\omega}{\sqrt{2(S_{q}-S)}}\right),
\end{eqnarray}

\noindent with $S_{q}=S(M_{\rm{EPS}}/q)$ and $S=S(M_{\rm{EPS}})$. Using the fact that $\lim_{x\rightarrow 0}\rm{erf}(x)\rightarrow 2x/\sqrt{\pi}$, yields

\begin{equation}\label{dMdo}
\frac{dM_{\rm{EPS}}}{d\omega}=-\sqrt{\frac{2}{\pi}}\frac{M_{\rm{EPS}}}{\sqrt{S_{q}-S}}.
\end{equation}

The differential equation for $M_{\rm{EPS}}$ (eq.~\ref{dMdo}) can be written in terms of redshift by replacing $d\omega = d(\delta_{0}/D(z))$ (given by eq.~\ref{delta}),

\begin{eqnarray}\label{dMdo1}
\frac{dM_{\rm{EPS}}}{dz}&=& -\sqrt{\frac{2}{\pi}}\frac{M_{\rm{EPS}}}{\sqrt{S_{q}-S}}\frac{1.686\Omega_{\rm{m}}(z)^{0.0055}}{D(z)}\\\nonumber
&& \times\left[\frac{0.0055d\Omega_{\rm{m}}(z)/dz}{\Omega_{\rm{m}}(z)}-\frac{dD(z)/dz}{D(z)}\right].
\end{eqnarray}

\noindent The above equation simplifies since $0.0055(d\Omega_{\rm{m}}(z)/dz)/\Omega_{\rm{m}}(z)\sim 0$ and $\Omega_{\rm{m}}(z)^{0.0055}$ can well be approximated by $1$, then the rate of change gives

\begin{eqnarray}\label{dMdz}
\frac{dM_{\rm{EPS}}}{dz}&=& \sqrt{\frac{2}{\pi}}\frac{M_{\rm{EPS}}}{\sqrt{S_{q}-S}}\frac{1.686}{D(z)^{2}}\frac{dD(z)}{dz}.
\end{eqnarray}

\section{Step-by-step guide to compute halo mass histories}\label{MAH_Description}

\subsubsection{Analytic model based on EPS}

This appendix provides a step-by-step procedure that details how to calculate the halo mass histories using the analytic model presented in Section 2:

\begin{enumerate}
\item Calculate the linear power spectrum $P(k)$. In this work we use the approximation of \cite{Eisenstein}.
\item Perform the integral 

\begin{equation}
S(R)=\frac{1}{2\pi^{2}}\int_{0}^{\infty}P(k)\hat{W}^{2}(k;R)k^{2}dk,
\end{equation}

\noindent where $\hat{W}^{2}(k;R)$ is the Fourier transform of a top hat window function and $R$ defines $S$ in a sphere of mass $M=(4\pi/3)\rho_{\rm{m,0}}R^{3}$, where $\rho_{\rm{m,0}}$ is the mean background density today.
\item Given $M_{0}$, the halo mass today, calculate the mass history by first obtaining $\tilde{z}_{\rm{f}}$

\begin{equation}
\tilde{z}_{\rm{f}}=-0.0064(\log_{10}M_{0})^{2}+0.0237(\log_{10}M_{0})+1.8837
\end{equation}

\noindent and

\begin{equation}
q=4.137\tilde{z}^{-0.9476}_{\rm{f}}.
\end{equation}

\item Use the parameter $q$ to calculate $f(M_{0})$, the function that relates the power spectrum to the mass history through the mass variance S,

\begin{equation}
f(M_{0})=1/\sqrt{S(M_{0}/q)-S(M_{0})}.
\end{equation}

\item Finally, the mass history can be calculated as follows,

\begin{eqnarray}
M(z)&=&M_{0}(1+z)^{af(M_{0})}e^{-f(M_{0})z},\\
a&=&\left[1.686(2/\pi)^{1/2}\frac{dD}{dz}|_{z=0}+1\right],
\end{eqnarray}

\noindent where $dD/dz$ is the derivative of the linear growth factor, which can be computed by performing the integral

\begin{equation}
D(z)\propto H(z)\int_{z}^{\infty}\frac{1+z'}{H(z')^{3}}dz'.
\end{equation}

\noindent $D(z)$ is normalized to unity at the present.

\end{enumerate}

The above model is suitable for any adopted cosmology and halo mass range.

\end{document}